\documentclass[preprint,amsmath, amssymb, aps, prd, showpacs,superscriptaddress]{revtex4-2}

\usepackage[utf8]{inputenc}
\usepackage{hyperref}
\usepackage{todonotes}
\usepackage{physics}
\usepackage{mathtools}
\usepackage{dsfont}
\usepackage{bbm}
\usepackage{comment}
\usepackage{enumerate}
\usepackage{amsthm}

\newcommand{\A}{{\mathcal A}}
\newcommand{\Hcal}{{\mathcal H}}
\newcommand{\B}{{\mathcal B}}
\newcommand{\one}{{\mathds 1}}

\newcommand{\R}{{\mathbb R}}
\newcommand{\C}{{\mathbb C}}

\numberwithin{equation}{section}

\begin{document}

\title{Uncertainties in Quantum Measurements: A Quantum Tomography}

\author{A.P. \surname{Balachandran}}
\email{balachandran38@gmail.com}
\affiliation{Department of Physics, Syracuse University, Syracuse, New York 13244-1130, USA}

\author{F. \surname{Calder\'on}}
\email[]{fcalder@umich.edu}
\affiliation{Departament of Philosophy, University of Michigan, 435 South State Street 2215 Angell Hall, Ann Arbor, MI 48109-1003, USA}

\author{V.P. \surname{Nair}}
\email[]{vpnair@ccny.cuny.edu}
\affiliation{City College of the CUNY, New York, New York 10031, USA}

\author{Aleksandr  \surname{Pinzul}}
\email{aleksandr.pinzul@gmail.com}
\affiliation{Universidade de Brasília, Instituto de Física 70910-900, Brasília, DF, Brazil}
\affiliation{International Center of Physics C.P. 04667, Brasília, DF, Brazil}

\author{A.F. Reyes-Lega}
\email[]{anreyes@uniandes.edu.co}
\affiliation{Departamento de F\'{i}sica, Universidad de los Andes,  A.A. 4976-12340, Bogot\'a, Colombia}

\author{S. \surname{Vaidya}}
\email[]{vaidya@iisc.ac.in}
\affiliation{Centre for High Energy Physics,  Indian Institute of Science, Bengaluru, 560012, India}

\begin{abstract}
The observables associated with a quantum system $S$ form a non-commutative algebra ${\mathcal A}_S$. It is assumed that a density matrix $\rho$ can be determined from the expectation values of observables. But $\mathcal A_S$ admits inner automorphisms $a\mapsto uau^{-1},\; a,u\in {\mathcal A}_S$, $u^*u=u^*u=\one$, so that its individual elements can be identified only up to unitary transformations. So since $\mathrm{Tr} \rho (uau^*)= \mathrm{Tr} (u^*\rho u)a$, only the spectrum of $\rho$, or its characteristic polynomial, can be determined in quantum mechanics. In local quantum field theory, $\rho$ cannot be determined at all, as we shall explain. However, abelian algebras do not have inner automorphisms, so the measurement apparatus can determine mean values of observables in abelian algebras ${\mathcal A}_M\subset {\mathcal A}_S$ ($M$ for measurement, $S$ for system). We study the uncertainties in extending $\rho|_{{\mathcal A}_M}$ to $\rho|_{{\mathcal A}_S}$ (the determination of which means measurement of ${\mathcal A}_S$) and devise a protocol to determine
$\rho|_{{\mathcal A}_S}\equiv \rho$ by determining $\rho|_{\A_M}$ for different choices of $\A_M$. The problem we formulate and study is a generalization of the Kadison-Singer theorem. We give an example  where the system $S$ is a particle on a circle and the experiment measures the abelian algebra of  a magnetic field $B$ coupled to $S$. The  measurement of $B$ gives information about the state $\rho$ of the system $S$ due to operator mixing. Associated uncertainty principles for von Neumann entropy are discussed in the appendix, adapting the  earlier work of Bia{\l}ynicki-Birula and Mycielski~\cite{BialynickiBirula1975} to the present case.
\end{abstract}

\maketitle
\tableofcontents
\section{Introduction}
Ever since Dirac introduced the notion of ``complete commuting sets of observables'' (CCS), abelian algebras are routinely used in quantum physics for labeling basis vector states,\footnote{We use the term ``vector state'' for both a state on $\A_S$ and a ray in a background Hilbert space $\Hcal$.} both in quantum mechanics and in quantum field theory. There have also been several publications on the appearance of commutative algebras in measurements, notably by Hepp \cite{Hepp1972}, Araki and Yanase \cite{Araki1960}, Fioroni and Immirzi \cite{Fioroni1994}, and Wightman \cite{Wightman1995}. Significant is the work of Klaas Landsman \cite{Landsman2017a,Landsman2017}, who has argued that the classical/quantum distinction can be adapted, with a few caveats, to that of commutative/non-commutative algebras. Some of the issues motivating the use of C$^*$-algebras as a unified language for both classical and quantum observables stem from the early days of quantum physics, in particular from the Einstein-Bohr debate \cite{Howard2007,Landsman2006}, as Landsman's notion of ``Bohrification" illustrates. His approach distinguishes studying commutative C$^*$-subalgebras of non-commutative C$^*$-algebras and extending commutative C$^*$-algebras to non-commutative ones. The first strategy is particularly relevant for this paper since we want to address the following problem posed by Kadison and Singer \cite{Kadison1959}: when the operators of a CCS have a discrete spectrum, does a pure state on said CCS extend uniquely and as a pure state to the full algebra of observables? They conjectured in 1959 that this is indeed the case when $\Hcal$ is separable and $\A_S$, the (in general non-commutative) algebra of observables of a quantum system, is equal to $\B (\Hcal)$, the algebra of all bounded operators on the Hilbert space $\Hcal$. The conjecture was established as a theorem only in 2015 by Marcus, Spielman, and Srivastava \cite{Marcus2015}. The difficulties in the proof are caused by the so-called non-principal ``normal'' states on $\B(\Hcal)$, which can only be proved to exist  using the axiom of choice.
These states, apparently, seem irrelevant for physics~\cite{Marcus2017}.
We note that the Kadison-Singer theorem does not extend to the case where a CCS contains operators with continuous spectra. The corresponding vector states,
routinely used in many contexts in physics, are not normalizable and hence do not belong to $\Hcal$.

The problems we address in this paper center around the possibility of using the expectation values of the abelian subalgebras of $\A_S$ for the case where $\A_S$ is non-commutative. This possibility will be explained in more detail in section \ref{sec:2}. The ambiguities or uncertainties that arise in this process are
present both for a simple qubit and local algebras of algebraic quantum field theory. If we measure just one abelian subalgebra $\A_M\subset\A_S$, the best we can do is to determine a state $\rho$ \textit{when restricted to $\A_M$}, the latter being generated by the CCS, and then study its extensions to $\A_S$. As already mentioned, when
$\rho|_{\A_M}$ is pure, the extension is unique and pure, and that is the lucky case. However, this is not the case if $\rho|_{\A_M}$ is not pure. It is in this sense that our paper goes along the same line as the Kadison-Singer theorem.

A simple quantum tomographic protocol for implementing the recovery of $\rho|_{\A_S}$ is discussed in section \ref{sec:3}, in terms of
measurements of $\rho|_{u\A_Mu^*}, u\in \A_S$, where $u^*u=\one=uu^*$.
(We will use ``$*$''
to denote the hermitian adjoint henceforth.) Thus one measures the restriction of $\rho$ to various
abelian subalgebras $u\A_Mu^*$ which are $*$-isomorphic. We then show how $\rho$ can be uniquely recovered as a state on $\A_S$
by Fourier transform on the group of $u's$, or rather on its orbit
of $\A_M$. We remark that this orbit is the real Grassmannian $Gr(1, \R_N)$.
This tomographic procedure works fine if $\A_M$ is a finite-dimensional algebra such as $\B_N(\Hcal)$ and perhaps in its limit $\B(\Hcal)$ for large $N$. In short: the complete determination of $\rho$ requires measurements of different commutative subalgebras, generated by automorphisms on a particular abelian subalgebra we will specify below. Actually, we can do better: it is enough to measure finitely many abelian subalgebras to recover the full state, as we will explain. It is uncertain whether this algorithm can be extended to the local algebras of algebraic quantum field theory, which are hyperfinite type $\mathrm{III}_1$ von Neumann factors. We will discuss some difficulties pointing in that direction.
Our method of recovering $\rho$ from the Grassmannian is not new. It is due to Man'ko and Man'ko \cite{Manko2011}. We will explain the connection between their work and ours later. However, our motivations differ from theirs.
We also mention that there are several notable papers on quantum tomography by Alberto Ibort, Margarita and Vladimir Man'ko, Giuseppe Marmo, and Franco Ventriglia (cf.~\cite{Manko2004,Ibort2009} and references therein).

Other material in this paper concerns the determination of a state of a particle on a circle by coupling it to a magnetic field. In the appendix we discuss entropic inequalities for Fourier transforms on Lie groups, adapting the
known inequalities  for quantum mechanical systems and Fourier transforms on $\R^N$.

\section{Ambiguities in the determination of a state on $\A_S$}
\label{sec:2}
The algebra $\A_S$ is a non-abelian $*$-algebra with identity $\one$. We use the term ``state'', $\omega_\rho$, on $\A_S$ in the usual sense, as a positive linear functional on $\A_S$ which gives $1$ on the identity of $\A_S$. We also require the state to be normal so that mean values $\omega_\rho(a)$ of observables $a$ are given by a density matrix $\rho$, a non-negative trace 1 operator (we always assume that the relevant Hilbert space, $\Hcal$, is separable):
\begin{equation}
\omega_\rho(a) = \Tr (\rho a),\quad a\in \A_S.
\end{equation}
Henceforth we will identify $\omega_\rho$ with $\rho$ and refer to the latter as a state.

We start with the presentation of the basic problem we are addressing.
We will present the discussion for now in the finite-dimensional case, but it can probably be adapted to infinite dimensions, so long as we stay within $\B(\Hcal)$, the algebra of ``quasi-local'' observables such as those in scattering theory in the absence of superselection sectors. However, local algebras are type $\mathrm{III}_1$ factors and require special considerations.

Let us consider the situation where $\A_S$ is presented abstractly, say in terms of generators $a_i$ and relations among them as
\begin{equation}
\label{eq:2.2}
\A_S =\langle a_i, i=1,2,\ldots : \quad a_i a_j =N_{i j}^k a_k \rangle.
\end{equation}
$\A_S$ is typically non-abelian so that $N_{i j}^k\neq N_{ji}^k$ in general. For that reason, there exist non-trivial inner automorphisms, i.e., there are operators $a_i^u$ unitarily related to $a_i$ for $u\in\A_S$, and they too generate $\A_S$ and have the same relations:
\begin{equation}
\label{eq:2.3}
a_i^u= u a_i u^*, \quad uu^*=u^* u =\one, \quad a_i^u a_j^u =N_{i j}^k a_k^u.
\end{equation}
There is no way to identify a particular $a_i\in \A_S$ from (\ref{eq:2.2}): there is always the ambiguity of (\ref{eq:2.3}).
Since
\begin{equation}
\Tr (\rho a_i^u)= \Tr (u^*\rho u a_i),
\end{equation}
two different experimentalists seeking to identify the state from its mean values for observables in $\A_S$ can determine it only up to unitary transformations, unless some \emph{external} labeling of the elements of $\mathcal A_S$ has been done (see also below). This means that only its spectrum or characteristic polynomial can be determined, knowing just (\ref{eq:2.2}). This is the case for $\B_N(\Hcal)$ or $\B(\Hcal)$. We will come to the local algebras below.

One option which seemingly can overcome this ambiguity is to couple $\A_S$ to a reference algebra $\A_M$, which may be external or a subalgebra of $\A_S$. If $\A_M$ is external, we can consider $\overline{\A_S\vee \A_M}$ and take this as the relevant $\A_S$, but now $\A_M \subset \A_S$.
In other words, a reference algebra $\A_M \subset \A_S$ covers the general case. However,
if $\A_M$ is also non-abelian, the same problem is encountered for measurements of $\A_M$, leading to a continued recursion rather than resolving the problem.
This motivates us to assume the point of view that the measured algebra $\A_M \subset \A_S$ must be abelian to go towards a determination of $\rho_S\equiv\rho$.
This conclusion is acceptable, as abelian algebras determine classical systems, and $\rho|_{\A_M}$ is a classical probability distribution satisfying Kolmogorov's axioms. In such a situation, we may note that,
happily, there is also no Schrödinger's cat paradox for $\A_M$.

However, once we are at this point, the Kadison-Singer question becomes the necessary next step that one must address: How do we extend $\rho|_{\A_M}$ to the full $\rho_S$ on $\A_S$ and what are the attendant ambiguities? Any experiment will have to confront this issue. We will take this up in section \ref{sec:3}.

Turning to the local case, recall that the local algebras of observables $\A_S$ of algebraic quantum field theory are (hyperfinite) type $\mathrm{III}_1$ von Neumann factors. Here we are confronted by the profound theorem of Connes and St{\o}rmer \cite{Connes1978}, which proves that, given two normal states $\rho$ and $\sigma$ on $\A_S$ and any $\varepsilon > 0$, there exists a unitary
element $u\in \A_S$ such that
\begin{equation}
\label{ec:p14.1}
\| u^* \rho u -\sigma\|_1 < \varepsilon.
\end{equation}
See  (\ref{eq:distance}) for the definition of  $\|\cdot\|_1$.
The theorem shows that this property is necessary and sufficient to characterize a von Neumann factor of type $\mathrm{III}_1$.
Given that individual elements of $\A_S$ can be determined only up to inner automorphisms, it is clear that it is impossible to distinguish two normal states by measurements on a local algebra. Compare this with the quantum mechanical (matrix) case we addressed before: the Kadison-Singer ambiguity is due to the \emph{extension}, but in this case we have problems determining states from the outset.
Note that if the algebra is of any other type than $\mathrm{III}_1$ we do have to some extent the possibility to distinguish between the states on the algebra by purely intrinsic means. The point is that, in cases other than $\mathrm{III}_1$, the space of equivalence classes (under the action of unitaries from $\mathcal A_S$) is non-trivial, and the (metric) space of the orbits will have a finite size. For example, it will be equal to $2(1-1/n)$ for  type $I_n$ algebras. Clearly, this will allow to distinguish between the states belonging to different orbits (in contrast to the type $\mathrm{III}_1$ case). See \cite{Connes1985} for the details.

The Connes-St{\o}rmer theorem is applicable for non-gauge quantum field theories with a mass gap (to avoid infrared problems). For such theories,
a natural question is: how do we see the emergence of the CCS accessible to experiments? In an interesting paper, Fioroni and Immirzi \cite{Fioroni1994} have argued that the measurable operators emerge from the labels of superselection sectors. And, since these labels take constant values in each sector, they are simultaneously diagonal and hence commute. They have also argued that the measuring apparatus should be in an unstable quantum state, which then undergoes a first-order phase transition to
(perhaps a mixture of) superselection sectors because of the disturbance
caused by the switching on of its coupling with $\A_S$ for observations. See also Wightman \cite{Wightman1995}.

Concrete models for the emergence of commutative algebras from time evolution have also been constructed by Hepp \cite{Hepp1972}.
Hepp's work is based on the concept of ``observables at infinity'' by Lanford III and Ruelle \cite{Lanford1969} and can  probably be adapted to the phase transitions of Fioroni and Immirzi \cite{Fioroni1994}.

In gauge theories, including QED, many superselection sectors arise from infrared effects and are physically natural. Whether a gauge theory should be an essential underlying feature for all physical measurements is an intriguing question. We will return to gauge theories elsewhere.

\section{A tomography to determine $\rho_S$ from $\A_M \subset \B_N(\Hcal)$}
\label{sec:3}

An affirmative answer to the feasibility of identifying $\rho_S$ by measurements on different abelian subalgebras $\A_M$ has still to be identified.
We need to know how this can be carried out in practice.
In other words, a protocol for the reconstruction of $\rho_S$ from
$\rho|_{\A_M}$ is important. We now turn to this issue.

Our reconstruction below of $\rho_S$ from
$\rho|_{\A_M}$ applies only to $\B_N(\Hcal)$
or the quasi-local algebra $\B(\Hcal)$. We  do not need to consider superselection sectors in either case since they have only one irreducible $*$-representation, up to unitary $*$-equivalence.

The density matrix $\rho$ is a positive-definite trace 1 matrix. We will focus on $\B_N$, in which case it is an $N\times N$ matrix.
By definition, $\rho$ is of rank $k$ if the image of $\C^N$
under $\rho$ is $\C^k$. In this case, $k$ rows of $\rho$ (or, equivalently, $k$ columns of $\rho$) are linearly independent.

The evaluation of $\rho$ as a state on the maximal abelian subalgebra\footnote{Note that at this step we are already making some preferred choice of $\mathcal A_M$. Namely, we start with the diagonal maximal abelian subalgebra. This sets up some kind of a ``reference frame'' and all other abelian subalgebras will be defined with respect to this choice. See also the futher discussion.}
\begin{equation}
\A_D:=\left\lbrace A_D= \mbox{$\sum_m$} a_m P_m \in \B_N(\Hcal):\; a_m\in \C,\; (P_m)_{ij}=\delta_{i,j}\delta_{i,m}
\right\rbrace
\end{equation}
gives
\begin{equation}
\Tr \rho A_D = \sum_ma_m \rho_{m,m}
\end{equation}
which, since $a_m$ are known and can be chosen to be any element of $\C$, fixes the diagonal elements
of $\rho$. Thus measurement of $\A_D$ gives the diagonal map (we denote $\rho_D:=\rho|_{\mathcal A_D}$)
\begin{equation}
\rho \mapsto \rho_D,\quad (\rho_D)_{m,m}=(\rho)_{m,m}, \quad (\rho_D)_{i,j}=0\; \mbox{ if } \;i\neq j.
\end{equation}
For example, for $N=2$, the general form of $\rho_M$ is
\begin{equation}
\label{ec:condition}
\rho=\left(
\begin{array}{cc}
\lambda_1 & \alpha \\
\bar \alpha & \lambda_2 \\
\end{array}
\right),\quad \lambda_i\ge 0,\;\sum_i\lambda_i=1,\;
\alpha\in \C,\;\lambda_1\lambda_2-|\alpha|^2 \ge 0,
\end{equation}
while
\begin{equation}
\rho_D =\left(
\begin{array}{cc}
\lambda_1 & 0 \\
0 & \lambda_2 \\
\end{array}
\right).
\end{equation}
Thus it remains to determine $\alpha$. At this point, our previous comment on the impossibility to determine the whole state $\rho$
from performing measurements only on a diagonal subalgebra should be clear. There is a family of ``lifts'' of a given state $\rho_D$ to the full state on $\mathcal B_N(\Hcal)$. In the $2\times 2$ case, it is parameterized by $\alpha\in\C$ satisfying the condition in (\ref{ec:condition}). There is a natural way to characterize the emergent uncertainty. If one defines a distance between two states, $\rho_1$ and $\rho_2$, by
\begin{equation}
\label{eq:distance}
d_{\rho_1,\rho_2}:= \|\rho_1-\rho_2\|_1\equiv \underset{a \in \mathcal B_N(\Hcal), \|a\|\le 1}{\mathrm{sup}}\Tr|\rho_1 a-\rho_2 a|,
\end{equation}
(this is the distance in (\ref{ec:p14.1})) then it is natural to define the uncertainty $\Delta \rho_D $ as the maximum distance between two lifts of the same diagonal state, $\rho_D$,
\begin{equation}
\Delta \rho_D := \underset{\rho', \rho''\in L_D}{\mathrm{sup}}
\|\rho'-\rho''\|_1,
\end{equation}
where $L_D$ denotes the set of all lifts of $\rho_D$. In the $2\times 2$ case it is not difficult to calculate this explicitly, the result being
\begin{equation}
\Delta \rho_{D}= 2\sqrt{\lambda_1\lambda_2}.
\end{equation}
So, if the original state was pure ($\lambda_1=0$ or $\lambda_2=0$), we would have $\Delta \rho_{D}=0$, in agreement with the Kadison-Singer theorem.

Our strategy to determine $\rho$ is to measure it on all abelian subalgebras $\A_D$ $*$-isomorphic to $\A_D$. They are obtained by unitary transformations of $\A_D$:
\begin{equation}
\label{eq:3.6Au}
\A_D^u= \lbrace A^u=\sum_\ell a_\ell uP_\ell u^*,\; uu^*=u^* u = \one\rbrace.
\end{equation}
It turns out to be sufficient to measure $\rho$ on the orbit under $U(N)$ of one rank 1 projector, say $P_1$.
This can be shown explicitly as follows. Notice that
$u$ is an element of the $N\times N$ irreducible representation of the group $U(N)$. The CCS labelling the basis of $\A_S$ is a Cartan subalgebra of $\frak u(N)$, with basis  $h_i\; (i=1,\ldots,N-1)$ and $h_0=\sqrt{\frac{2}{N}}\one$. They have the canonical normalization
\begin{equation}
\Tr (h_i h_j) = 2\delta_{ij}
\end{equation}
and of course $\Tr h_i=0$ if $i \ge 1$.

Now consider the orbit of $P_1$ under the action of $u$, given by
$u P_1 u^* = u P_1 u^{-1}$, namely,
\begin{equation}
\label{eq:3.8}
(u P_1 u^*)_{\alpha\beta} = u_{\alpha 1}u^*_{1 \beta}.
\end{equation}
Notice that $\mathrm{Ad}(U(N))P_1=U(N)/U_{P_1}= \mathrm{Ad}(SU(N))P_1$,
where $U_{P_1}$ is a stability subgroup. Hence, elements of the form $e^{i\varphi h_0}$, being overall phases, cancel out in (\ref{eq:3.8}).

But there is more that we can factor out. Write
\begin{equation}
P_1 = \sum_i p_i h_i, \quad p_i\in \R.
\end{equation}
It is convenient to choose $h_i$ so that
\begin{equation}
h_1 = \left[\frac{2N}{N-1}\right]^{1/2} \left( P_1-\frac{1}{N}\one\right).
\end{equation}
Then if $E_k$ is an orthogonal basis for the Lie algebra of $SU(N)$ with the normalization $\Tr E_kE_\ell=2\delta_{k\ell}$ and with $E_1=h_1$, we get
\begin{equation}
uP_1u^{-1}=\left[\frac{N-1}{2N}\right]^{1/2} E_k D_{k1}(u)+
\frac{1}{N} \one,
\end{equation}
where $D(u)$ is the adjoint representation of $u$. Thus if $\mbox{Ad}SU(N-1)$ is the subgroup of $\mbox{Ad} SU(U_N)$ with elements
of the form
\begin{equation}
\left(
\begin{array}{c|ccc}
1 & 0 \quad 0 & \cdots & \\
\hline\\
0 & &(N-1) \times (N-1) & \\
0 & & \mbox{Ad} SU(N-1) & \\
\vdots & & & \\
\end{array}
\right),
\end{equation}
its action on the right of $u$ does not affect the point of the orbit. The orbit is thus the real Grassmannian
\begin{equation}
\label{ec:Grassmannian}
\mbox{Ad} SU(N)/ \mbox{Ad} SU(N-1)= \mbox{Gr}_\R(1,N),
\end{equation}
which is the space of lines through the origin in $\R^N$.
The reality is due to the fact that $\mbox{Ad}SU(N)$ is a real representation.

Let
\begin{equation}
\rho_k:= \Tr (\rho E_k).
\end{equation}
Since $\one$ and $E_k$ form a basis for $\mbox{Mat}_N(\C)$, and $\Tr\rho\one =1$, we can fully reconstruct $\rho$ from knowing $\rho_k$. We can do that from the evaluation of $\rho$ on $uP_1u^{-1}$. (This reconstruction is identical to the one due to M. Manko and V. Manko \cite{Manko2011} as we will also show.)
Now
\begin{equation}
\Tr \rho(u P_1u^{-1})=\frac{1}{N} +
\left[\frac{N-1}{2N}\right]^{1/2} \rho_k D(u)_{k1}.
\end{equation}
This can be inverted for $\rho_k$ in a straightforward way
using group orthogonality. If $d\mu$ is the invariant measure for a compact Lie group $G$ and $D^{(\rho)}, D^{(\sigma)}$ are two of its unitary irreducible representations, then we have the orthogonality relation
\begin{equation}
\label{eq:orthogonality}
\int_G d\mu(g) D_{\alpha\beta}^{(\rho) *}(g) D_{\gamma\lambda}^{(\sigma)}(g) = \frac{1}{d_\sigma}\delta_{\rho \sigma} \delta_{\alpha\lambda}\delta_{\beta \gamma},
\end{equation}
where $d_\sigma$ denotes the dimension of the representation
$D^{(\sigma)}$ and we have normalized the Haar measure as
\begin{equation}
\int_G d\mu (g)=1.
\end{equation}
For the adjoint representation, $d_\sigma=N^2-1.$
Hence
\begin{equation}
\int_G d\mu (u) D(u)^*_{\alpha\beta}\Tr\rho(u P_1 u^{-1})=
\frac{1}{(N+1)(2N(N-1))^{1/2}} \rho_\beta \delta_{\alpha 1}.
\end{equation}
This determines $\rho$ fully.

Although integration over $u$ does yield $\rho$, a finite number of
$u$'s suffices if they are judiciously chosen. This can be seen as follows.
The basis $\{ E_k \}$ for $\mathfrak{u}(N)$ consists of
$\{ h_i \}$, which is the basis for the Cartan subalgebra $h$,
and $\{ {\tilde E}_k\}$ which are the roots.
The measurements yield $\rho$ evaluated on the Cartan subalgebra.
We need $\rho$ evaluated on the roots as well to complete the set of
$\rho_k$.
This can be done by a suitable ``rotation" of the initial Cartan subalgebra.
Consider a particular root ${\tilde E}_k$. The adjoint orbit of ${\tilde E}_k$
(under the action of $SU(N)$) intersects $h$ in exactly one point. In other words, there is a $u_k \in SU(N)$ such that
$u^*_k {\tilde E}_k u_k \in h$. Equivalently,
$u_k h u^*_k$ is another Cartan subalgebra such that
${\tilde E}_k \in u_k h u^*_k$. Evaluation of $\rho$ (via another measurement)
leads to $\rho_k$. Since we have $N (N-1)$ roots, with the suitable choice of
$N(N-1)$ such $u$'s, one can completely determine $\rho$.

Note that, in this construction, we assumed that we measure a state on the whole abelian subalgebra $h$. But we can also recover $\rho$ by measuring all $N$ pairwise orthogonal rank one projectors $P_1$ using just $N^2-1$ unitary transformations.
Namely, take  $P_1\in h$ as before. Then, it is trivial to see that $u_k:=\sum_{1\leq s\leq N} P_{s,s+k-1}$ will map $P_1$ to $P_k\in h$. Here, $P_{k,s}$ is a partial isometry mapping the $k^{\mathrm th}$ basis vector to the $s^{\mathrm th}$ one and $s+k-1$ is $\mathrm{mod}\, N$. So, in this case, we need $N(N-1)+N-1=N^2-1$ unitary rotations. If we include the first measurement on $P_1$, we will get $N^2$ measurements, as one would expect for a system with $N^2$ degrees of freedom.

Another observation which is worthy of remark is about the relevance of
stochastic maps in the context of measurements of different abelian subalgebras.
The evaluation of the diagonal density matrix
$\rho_D (\lambda )= \sum_s \lambda_s P_s$ on the abelian algebras $u (\sum_s a_s P_s)u^*$ is equivalent to measuring $\rho_D (T \lambda )$,
where
\begin{equation}
\label{eq:3.19}
(T\lambda)_r= \sum_s T_{sr} \lambda_s, \quad T= (|u_{sr}|^2),\quad
\lambda = (\lambda_1,\lambda_2,\ldots,\lambda_N)).
\end{equation}
The key point is that $T$ is a doubly stochastic matrix since
\begin{equation}
T_{sr}\ge 0,\quad \sum_r T_{sr}=\sum_s T_{sr} = 1.
\end{equation}
It is straightforward to obtain the result
(\ref{eq:3.19}). We have
\begin{align}
\Tr \rho_D(\lambda)(u (\mbox{$\sum_r$} a_r P_r)u^*) &=
\sum_{s,r} \lambda_s a_r u_{sr}u^*_{rs}
= \sum_{s, r} |u_{sr}|^2 \lambda_s\, a_r \nonumber\\
&=\sum_{r} (T \lambda )_r a_r\nonumber\\
&= \Tr \rho_D(T\lambda) (\sum_r a_r P_r).
\end{align}
This shows the result. Note that entropy, being concave, is non-decreasing under the stochastic maps $T$.

It is also worth remarking on a couple of known structural results on stochastic matrices $T$. First, they form a compact convex set. Second, by a theorem of Birkhoff and von Neumann \cite{Bapat1997}, their extremal points are permutation matrices of dimension $N^2$, and every $\rho$ is a convex combination of at most $N^2 -2 N+2$ such matrices \cite{Marcus1959}. Therefore, the further study of the relevance of doubly stochastic matrices for our measuring protocol might be beneficial.

We now turn to the question of relating our considerations to the work of
Man'ko and Man'ko. Their excellent review article \cite{Manko2011} and,
in particular, their
equation (27) give the basis for the comparison.
They define the tomogram of $\rho$ as
\begin{equation}
W(m,u)= \langle m |u\rho u^*|m\rangle, \quad m= -\frac{(N-1)}{2},-\frac{(N-1)}{2}+1,\ldots,\frac{(N-1)}{2},\; u\in U(N)
\end{equation}
and discuss the reconstruction of $\rho$ from $W(m,u)$.
To conform to their notation, let $
P_m = |m\rangle\langle m|$
be a rank one projector. Then
\begin{equation}
W(m,u) = \Tr \rho (u^* P_m u).
\end{equation}
But this is our equation (\ref{eq:3.8})
for $m=1$ and with our $u$ replaced by $u^*$. They also consider the case where $u$ is the spin $j$ representation of the $SU(2)$ group. Since the crucial group orthogonality (\ref{eq:orthogonality}) is valid for $SU(2)$, we can see that our approach is not significantly different from theirs (see also section \ref{sec:5}). However,  the motivation is different.

\section{A model for quantum tomography}
\label{sec:4}
The previous section can be thought of as dealing with the measurements
on the algebra of observables by the apparatus. In other words, it pertains to
the process of registering the outcomes of measurements
by an apparatus. For a complete description, we need to couple the measurement apparatus $M$ to the system $S$ and outline how to recover the pre-coupling state $\rho$  using both  the system and the apparatus.
We now suggest a method to do so. It is an amalgamation of much of the previous work we have cited.

We will assume that $S$ and $M$ are decoupled at times $t<t_0$, so that the state $\rho$ is a tensor product $\rho_S(t)\otimes \rho_M(t)$ which might be time-dependent. During these times, it evolves by the Hamiltonian
of the system plus apparatus, which is of the form $H=H_S\otimes \one +\one\otimes H_M$. It is assumed that we know these Hamiltonians and $\rho_M$, but not $\rho_S$. We want to determine $\rho_S$. To simplify the presentation, we will assume that $H$ is time-independent.

During the time interval $t_0< t <t_0+T$, we couple $S$ and $M$ through an interaction Hamiltonian $H_I(t)$. In quantum field theory, it should be a local coupling to avoid causality problems
like the ones of  Sorkin's protocol \cite{Sorkin1993}, but there is no such restriction in finite dimensions. We switch off the interaction $H_I(t)$ for $t>t_0+T$. The state $\rho_S(t)\otimes \rho_M(t)$ will then evolve by the unitary operator
\begin{align}
V(t,t_0) &= V_0(t, t_0) V_I(t,t_0),\quad t\ge t_0,\nonumber\\
V_0(t, t_0) &=e^{-i(t-t_0)H},\nonumber\\
V_I(t,t_0)&= \mathcal T\exp\left(-i\int_{t_0}^t d\tau H_I'(\tau)\right),\nonumber\\
H_I'(\tau) & = e^{iH\tau} H_I(\tau)e^{-iH\tau}.
\end{align}
For times $t>t_0+T$, $V(t,t_0 +T)$ again becomes $V_0(t, t_0+T)$.

It is assumed, as is natural, that the coupling $H_I(t)$ for $t_0<t<t_0+T$ is sufficiently strong that it leaves a significant imprint on the time-evolved state
\begin{equation}
\label{ec:state(t)}
\rho(t) = V(t,t_0) \left(\rho_S(t_0)\otimes \rho_M (t_0)\right)V(t,t_0)^*, \quad t>t_0 +T.
\end{equation}
This state generically will be entangled even though the state at time $t_0$ was not.

The experimentalist measures expectation values of the abelian algebras
\begin{equation}
(\one\otimes u) (\one \otimes \mbox {$\sum_i$}a_i P_i) (\one\otimes u^*)
\end{equation}
of the measurement observables for the state (\ref{ec:state(t)}). As in the last section, one gets the mean values
\begin{equation}
\Tr \rho(t)\left(\one\otimes u P_i u^*\right).
\end{equation}

Now we encounter a difference with the last section. Let $n$ and $N$ be the dimensions of $\Hcal_S$ and $\Hcal_M$. Then $\rho(t)\in \mbox{Mat}_{n N}(\C)$ while $P_i\in \mbox{Mat}_{N}(\C)$. So integrating over $\mbox{Ad}SU(N)$ and using $\Tr \rho(t)=1$, we can only extract
\begin{equation}
\label{ec:4.5}
\Tr \rho(t)(\one\otimes E_k),\quad E_k\in \mathfrak{su}(N).
\end{equation}
This determines the traceless part of the $N\times N$ submatrix of $\rho$, but as before, the trace is $1$ (from $\Tr \rho=1$). That is, only the restriction of $\rho$ to $\one\otimes \A_M$ gets determined.
The key point is thus to determine $\rho_S$ from the knowledge
of the evaluation of $\rho(t)$, for $t> t_0 +T$, on
$\one\otimes \A_M$.
Significantly, this will depend on $\rho_S$.
This can be seen more explicitly as follows.
One can expand $\rho_S(t_0) \otimes \rho_M(t_0)$ in a basis for
$U(nN)$ as
\begin{equation}
\rho_S(t_0) \otimes \rho_M(t_0) = \sum_{\alpha ,k} \rho_{S\alpha} (t_0) \, \rho_{Mk}(t_0)
\, e_\alpha\otimes E_k.
\label{ec:4.5a}
\end{equation}
(We include the $n\times n$ identity as $e_0 $, $N\times N$ identity as
$E_0$. Also $\rho_{S 0} (t_0) \, \rho_{M 0}(t_0)$ is fixed by
the trace of $\rho$.) The evaluation of $\rho (t)$ from
(\ref{ec:state(t)}) on $\one \otimes \A_M$ is of the form
\begin{equation}
\Tr \rho(t) (\one \otimes E_l) = \sum_{\alpha, k} \rho_{S \alpha}(t_0)\,
\rho_{M k}(t_0) \, D_{\alpha k, 0l}(V),
\label{ec:4.5b}
\end{equation}
where $D_{\alpha k, 0l}(V)$ denotes the adjoint representation
of $V$ (in $U(nN)$). This is the basic data obtained from the
measurement. If $N \geq n$ and there is sufficient freedom in choosing
$H_I$ (e.g., $H_I = \sum C_{\alpha k}\, e_\alpha \otimes E_k$ and the coefficients $C_{\alpha k}$ can be chosen freely),
one can invert (\ref{ec:4.5b}) to obtain $\rho_{S \alpha}(t_0)$ which determines
$\rho_{S}(t_0)$. In this sense,
a knowledge of (\ref{ec:4.5}) constitutes a measurement.
In reality, each apparatus is tied to one or a small number of
possible choices for $H_I$, so only limited information about
$\rho_{S}$ can be obtained in any experiment.
This analysis also makes clear that for a good experiment, we need $N$, the number of degrees of freedom of the apparatus to be large enough to cover the degrees of freedom of the system, and we
also need a good theory that controls the couplings $H_I$.

As $N\rightarrow\infty$, perhaps when the matrix elements of $u$ are fixed (weak convergence), asymptotic formulae may exist for $u$. It would be very interesting to examine such limits in the analysis above.
\section{An example: a particle on a circle}
The system in this example is a particle on a circle with the free Lagrangian
\begin{equation}
\mathcal{L}_S=\frac{1}{2}\dot\varphi^2,\qquad e^{i\varphi}\in S^1.
\end{equation}
Measurements are performed on a spatially constant magnetic field $B$ coupled to $S$ for a time interval $0\leq t \leq T$. For $t<0$, we take the Lagrangian $\mathcal L_M$ for $B$ to be
\begin{equation}
\mathcal L_M= \frac{1}{2}\dot B^2 -\frac{1}{2}B^2.
\end{equation}
This and other choices are for illustrative purposes.

The coupling of $S$ and $M$ is taken to be
\begin{equation}
\mathcal L_I = \lambda(t) \dot\varphi B, \quad \lambda(t)\in \mathbb R, \quad \lambda(t)=0\; \mbox{ for }\; t<0\; \mbox{ or }\; t>T.
\end{equation}
The state at time $t=0$ is assumed to be
\begin{equation}
\Omega = \Omega_S \otimes \Omega_M,
\end{equation}
where
\begin{equation}
\Omega_S=|n\rangle\langle n|,\quad \langle e^{i\varphi}|n\rangle=e^{i n \varphi}, \quad n\in \mathbb Z,
\end{equation}
and
\begin{equation}
\Omega_M=|0\rangle\langle 0|,
\end{equation}
$|0\rangle$ standing for the harmonic oscillator ground state.

For $t<0$, $M$ measures the abelian algebra with generator
\begin{equation}\label{eq:1xB}
\one \otimes \hat B,
\end{equation}
the hat denoting an operator.
At time $t>0$, the state becomes $\Omega^t$ and the mean value of $\one\otimes \hat B$ becomes
\begin{equation}\label{eq:V-7}
\Omega^t (\one \otimes \hat B) =\Omega\left[ e^{iHt} (\one \otimes \hat B) e^{-iHt}\right].
\end{equation}

The argument of $\Omega$ also generates an abelian algebra isomorphic to that of $\one\otimes \hat B$. We will now solve for
$ e^{iHt} (\one \otimes \hat B) e^{-iHt}$. Due to operator mixing, it involves operators of both $S$ and $M$. So its expectation value in $\Omega$ involves $\Omega_S$ evaluated on a system observable such as its momentum in this example. In this way we get information on $\Omega_S$.

We can find (\ref{eq:V-7}) by solving the equation of motion governing $\one\otimes \hat B := \one\otimes \hat B (0)$. We simplify notation notation by writing $\one\otimes \hat B(0)$ as $\hat B(0)$, with a similar notation elsewhere.

The momentum $\hat\pi$ conjugate to $\hat \varphi$ is
\begin{equation}
\hat \pi =\dot{\hat\varphi} + \lambda(t) \hat B,
\end{equation}
while that of $\hat B$ is
\begin{equation}
\hat P =\dot{\hat B}.
\end{equation}
The Hamiltonian $H_S$ for the system is
\begin{equation}\label{eq:V-10}
H_S= \frac{1}{2}(\hat \pi -\lambda(t) \hat B)^2,
\end{equation}
while the apparatus Hamiltonians is
\begin{eqnarray}
H_M &=&\frac{1}{2}\hat P^2 +\frac{1}{2}\hat B^2.
\end{eqnarray}
The equations of motion following from the total Hamiltonian $H=H_S + H_M$ are
\begin{equation}
\label{eq:V-14}
\frac{d^2\hat\varphi(t)}{dt^2} +\frac{d}{dt}(\lambda(t)\hat B(t)) = \frac{d}{dt}\hat \pi(t)=0,
\end{equation}
\begin{eqnarray}
\frac{d^2\hat B(t)}{dt^2} +\hat B(t) &=& \lambda(t) \dot{\hat \varphi}(t)\nonumber\\
                                &=& \lambda(t)\left[ \hat \pi(t)-\lambda(t)\hat B(t)\right]
\end{eqnarray}
or, using also (\ref{eq:V-14}),
\begin{equation}\label{eq:lambda^2}
\frac{d^2\hat B(t)}{dt^2} +(1+\lambda(t)^2)\hat B(t) = \lambda(t) \hat \pi(0).
\end{equation}

Eq. (\ref{eq:lambda^2}) can be analyzed in the following way. Consider the usual  (not operator) differential equation:
\begin{equation}
\label{eq:Sasha*}
\ddot u + (1+\lambda(t)^2)u = \lambda(t)
\end{equation}
and the corresponding homogeneous equation:
\begin{equation}
\label{eq:Sasha**}
\ddot u_0 + (1+\lambda(t)^2)u_0 = 0.
\end{equation}
Here $u$ and $u_0$ are usual functions. By the standard arguments (\ref{eq:Sasha**}) has two linearly independent solutions $u_1$ and $u_2$ that we will fix by the initial conditions at $t=0$:
\begin{equation}
\left\lbrace
  \begin{array}{cc}
    u_1(0) = 1,\quad & \dot u_1(0)=0 \\
    u_2(0) = 0,\quad & \dot u_2(0)=0 \\
  \end{array}
\right. .
\end{equation}
Note that the Wronskian is
\begin{equation}
W(t) : = u_1(t)\dot u_2(t)- \dot u_1(t) u_2(t) \equiv W(0) = 1
\end{equation}
and a partial solution to (\ref{eq:Sasha*}) is as usual given by
\begin{equation}
u_{\textrm{par}}(t) = -\int_0^t d\tau \lambda(\tau) u_2(\tau)u_1(t)
+\int_0^t d\tau \lambda(\tau) u_1(\tau)u_2(t).
\end{equation}
It is then obvious that we can write the corresponding partial solution of
(\ref{eq:lambda^2}) as
\begin{equation}
\hat B_{\textrm{par}}(t) = \hat \pi(0) u_{\textrm{par}}(t).
\end{equation}
Then the general solution to (\ref{eq:lambda^2}) subject to the initial conditions $\hat B (0) \equiv \hat B$ and $\dot{\hat B}(0) \equiv \hat P$
is given by
\begin{equation}
\hat B (t) = \hat \pi u_{\textrm{par}}(t) + \hat B u_1(t) +\hat P u_2(t).
\end{equation}
For $t>\tau$ this takes the form
\begin{equation}
\hat B (t) = (\hat B -\lambda_2 \hat \pi)u_1(t)+(\hat P +\lambda_1 \hat \pi)u_2(t),
\end{equation}
where
\begin{equation}
  \lambda_i := \int_0^T d\tau \lambda(\tau) u_i(\tau).
\end{equation}
So we see that the result does not have a very strong dependency on the actual form of the interaction, $\lambda(t)$. The main effect will be the same: The apparatus algebra generated by $\hat B$ and $\hat P$ is mixed with the system algebra. This will give the possibility to extract the information about $\Omega_S$ by measuring the abelian subalgebra generated by $\hat B(t)$ (as was explained in general in the previous section).
%
%

It is noteworthy that if $\mathcal D_0$ is the domain for the Hamiltonian (\ref{eq:V-10}), then it is isospectral to the Hamiltonian
\begin{equation}
H_S'=\frac{1}{2}\hat \pi^2
\end{equation}
with domain
\begin{equation}
\mathcal D_{\lambda(t)} = e^{-i\lambda(t) \hat \varphi\hat B}\mathcal D_0.
\end{equation}

The example of this section also illustrates that the determination of $\Omega_S$ involves theory. The measurement we have illustrated only partially determines $\Omega_S$, namely its restriction to  (\ref{eq:1xB}). This is to be expected: measurements of more observables reveals more about $\Omega_S$, a feature reflected in actual experiments.

\section{Conclusions}
We have studied the uncertainties in the determination of a state from its restriction to an abelian subalgebra.  A protocol to determine the state from measurements of different abelian subalgebras has been proposed and illustrated by means of an explicit example, where a particle on a circle in a circle is  coupled to a magnetic field. The reconstruction of the states discussed here resembles a Fourier transform and lead naturally to entropic inequalities which are obtained as generalizations (from $\mathbb R^N$ to Lie groups) of known ones.
\section*{Appendix}
\label{sec:5}
The reconstruction we outlined involved a transition from
\begin{equation}
\label{eq:5.1/rho}
\hat \rho(u)=\sum_k \Tr (\rho E_k) D_{k1}(u)
\end{equation}
to $\Tr (\rho E_k)$ and hence to $\rho$. This may be regarded as a transition from a function $\tilde \rho$ on $U(N)$ (or rather the Grassmannian $\mbox{Gr}_\R(1,N)$) to the Fourier coefficients $\Tr(\rho E_k)$. This is analogous to the transformation of a wave function $\psi$ on $\R^N$ to its momentum space function $\tilde \psi$ in quantum theory.

Now for the latter, there are a number of entropic inequalities connecting
\begin{equation}
\langle \ln|\psi(x)|^2\rangle :=\int d^N x |\psi(x)|^2 \ln|\psi(x)|^2
\quad\mbox{and} \quad
\langle \ln|\tilde\psi(p)|^2\rangle :=\int d^N p |\tilde\psi(p)|^2\ln|\tilde\psi(p)|^2,
\end{equation}
with the interpretation of $\psi$ and $\tilde \psi$ as configuration and momentum space wave functions. They are discussed in~\cite{BialynickiBirula1975} and elsewhere. We will comment on them below as well. It is natural to wonder what their analogues are for (\ref{eq:5.1/rho}).

Only one irreducible representation (of $SU(N)$) occurs in (\ref{eq:5.1/rho}). So let us consider the case where several irreducible representations occur, following a remark by
M. Manko and V. Manko~\cite{Manko2011}. For this purpose, let us replace $SU(N)$ by the spin $J=(N-1)/2$ unitary irreducible representation of $SU(2)$. Under the adjoint action of this $SU(2)$, the projectors $P_i$ will split into the direct sum of matrices $\Lambda_m^j$, $m\in \lbrace-j,-j+1,\ldots,j\rbrace$, $j=0,1,\ldots, 2J$. We can write
\begin{equation}
P_i = \sum_{m,i} c_{i m}^j\Lambda_m^j,
\end{equation}
where
\begin{equation}
[J_3, \Lambda_m^j] = i m \Lambda_m^j, \Tr (\Lambda_m^j\Lambda_{m'}^{j'})=
2\delta_{j,j'}\delta_{mm'},
\end{equation}
$J_3$ being the third component of angular momentum. Then, if $U$ is the unitary representation of $SU(2)$,
\begin{equation}
\label{eq:UPU}
U(g)P_i U(g)^{-1} = \sum_{m,m',j} c_{i m}^j \Lambda_{m'}^j
D_{m'm}^j(g),\quad g\in SU(2).
\end{equation}
Thus $U(g)$ replaces the $u$ in (\ref{eq:3.6Au}) while $D^j$ is the angular momentum $j$ representation of $SU(2)$. The new abelian algebra is
\begin{equation}
\A_D^{U(g)}=U(g) \A_D U(g)^{-1}.
\end{equation}
The expectation values of (\ref{eq:UPU}) are
\begin{equation}
\label{eq:5.7}
\Tr \rho (U(g)P_i U(g^{-1})) = \sum c_{i m}^j
\left(\Tr \rho \Lambda_{m'}^j\right)D_{m'm}^j(g).
\end{equation}
This equation clearly shows the ``Fourier'' expansion in rotation matrices: the Fourier coefficients are
\begin{equation}
\label{eq:5.8-Fourier-coeff}
\Tr \rho \Lambda_{m'}^j = \rho_{m'}^j.
\end{equation}
From a knowledge of (\ref{eq:5.8-Fourier-coeff}), we can write down $\rho$. But we can find $\rho_{m'}^j$ as before. With
\begin{equation}
\int_{g\in SU(2)} d\mu(g)=1,
\end{equation}
we get, using (\ref{eq:orthogonality}),
\begin{equation}
\int d\mu (g) D_{nn'}^\ell(g)^*\left( \Tr \rho(U(g)P_iU(g)^{-1})\right)
=\frac{1}{2\ell+1}c_{in}^\ell  \rho_{n'}^\ell.
\end{equation}
We are done.

What are the entropic inequalities governing (\ref{eq:5.7})?

The emergence of entropic inequalities from certain Banach spaces were first noted by Bia{\l}ynicki-Birula and Mycielski~\cite{BialynickiBirula1975}.

Thus consider the Banach space $L^p(\mathbb R^n)$ of functions $\Psi$ on $\mathbb R^n$ for $p>1$ and with norm
\begin{equation}
\|\psi\|_p = \left(\int d^n x |\psi(x)|^p\right)^{1/p}.
\end{equation}
The space dual to $L^p(\mathbb R^n)$ which gives the linear functionals on $L^p(\mathbb R^n)$ is $L^q(\mathbb R^n)$, where
\begin{equation}
\label{eq:1/p1/q}
\frac{1}{p}+\frac{1}{q}=1.
\end{equation}
Let us restrict $p$ to the interval $(1,2]$ so that $q\ge 2$:
\begin{equation}
\label{eq:5.13}
p\in (1,2]; \quad q\ge 2,
\end{equation}
and let $\tilde \psi$ be the Fourier transform of $\psi$:
\begin{equation}
\tilde \psi(k)=\frac{1}{(2\pi)^{n/2}}\int d^n x e^{-ik\cdot x} \psi(x).
\end{equation}
Then $\tilde \psi \in L^q(\mathbb R^n)$ and
\begin{equation}
\label{eq:5.14}
\kappa (p,q)\|\psi\|_p - \|\tilde \psi\|_q \ge 0,
\end{equation}
where
\begin{equation}
\kappa (p,q) =\left(\frac{2\pi}{q}\right)^{\frac{n}{2q}}\left(\frac{2\pi}{p}\right)^{-\frac{n}{2p}}
\end{equation}
and $q$ is determined by (\ref{eq:1/p1/q}) in terms of $p$.

At $q=2, p$ is also 2 and $\|\psi\|_2=\|\tilde\psi\|_2$ by the Parseval-Plancherel theorem. At $q=2$, therefore, L.H.S. of (\ref{eq:5.14}) is zero. So it cannot decrease as $q$ increases from 2 or the derivative of (\ref{eq:5.14}) on $q$ must be greater or equal than zero as $q$ approaches 2 from above. This gives the entropic inequality
\begin{multline}
\frac{n}{4}N(1+\ln\pi) -\frac{1}{2N}\int d^n x|\psi(x)|^2\ln|\psi(x)|^2 \\
-\frac{1}{2N}\int d^nk |\tilde \psi(k)|^2 \ln |\tilde \psi(k)|^2+ N\ln N\ge 0,
\end{multline}
where
\begin{equation}
N=\|\psi\|_2 = \|\tilde \psi\|_2.
\end{equation}
For normalized wave functions, $N=1$.

The inequality $\|\psi\|_p\ge \|\tilde\psi\|_q$ is known as the Hausdorff-Young (HY) inequality~\cite{Reed1975}. The determination of the precise coefficient $\kappa(p,q)$ came later and is due to Babenko~\cite{Babenko1961} and Beckner~\cite{Beckner1975}.

The HY inequality has been generalized  to Fourier transforms on groups and are discussed with references in~\cite{Cowling2019}. As an example, we reproduce the inequality for $U(1)$ reported in ~\cite{BialynickiBirula1975}.

Let $\Phi$ be a function  on $U(1)$. It has the Fourier expansion
\begin{equation}
\label{eq:Fourier-expansion}
\Phi(\varphi) = \sum_{m=-\infty}^\infty c_m e^{im\varphi}.
\end{equation}
Then with (\ref{eq:5.13}),
\begin{equation}
\label{eq:5.18B}
\|\Phi\|_p=\left(\int_0^{2\pi}\frac{d\varphi}{2\pi}|\Phi(\varphi)|^p\right)^{1/p}
\ge \left(\sum_{m=-\infty}^{\infty}|c_m|^q\right)^{1/q} \end{equation}
At $q=2$, the inequality is saturated. So again by differentiating with respect to $q$ and evaluating at $q=2$, we get the entropic inequality
\begin{equation}
-\int_0^{2\pi}\frac{d\varphi}{2\pi}|\Phi(\varphi)|^2\ln|\Phi(\varphi)|^2-
\sum_{m=-\infty}^{\infty}|c_m|^2\ln|c_m|^2\ge 0
\end{equation}
between canonically conjugate variables.

We now generalize this result to a generic compact connected Lie Group $G$. We can then adapt the result to (\ref{eq:5.7}).

From (\ref{eq:orthogonality}) we can get an orthonormal basis $d_{\alpha\beta}^\rho$ for $L^2(G)$:
\begin{equation}
d_{\alpha\beta}^\rho =\sqrt{d_\rho} D_{\alpha\beta}^\rho,
\end{equation}
\begin{equation}
\int_G d\mu (g) d_{\alpha \beta}^{\rho}(g)^*d_{\gamma \lambda}^\sigma(g)= \delta_{\rho\sigma} \delta_{\alpha\lambda} \delta_{\beta\gamma}.
\end{equation}
For a function $f$ on $G$, we can then write
\begin{equation}
f(g) =\sum_{\rho,\alpha,\beta} \hat f^\rho_{\alpha\beta} d^\rho_{\alpha\beta}.
\end{equation}
This enables us to introduce $L^p$ spaces for functions on $G$ and its Fourier coefficients, following (\ref{eq:Fourier-expansion}):
\begin{eqnarray}
\|f\|_p &:=& \left( \int_G d\mu(g) |f(g)|^p\right)^{1/p},\nonumber\\
\|\hat f\|_q &:=& \left( \sum_{\rho,\alpha,\beta}  |\hat f^\rho_{\alpha\beta}|^q\right)^{1/q}.
\end{eqnarray}
We the have the HY inequality
\begin{equation}
\|f\|_p \ge \|\hat f\|_q,\qquad q\ge 2, \;\frac{1}{p}+\frac{1}{q}=1,
\end{equation}
as in (\ref{eq:5.18B}) and the corresponding entropic inequality
\begin{equation}
-\int_G d\mu(g)|f(g)|^2\ln|f(g)|^2-
\sum_{\rho,\alpha,\beta}|\hat f^\rho_{\alpha,\beta}|^2\ln|f^\rho_{\alpha,\beta}|^2\ge 0.
\end{equation}
Identifying the LHS of (\ref{eq:5.7}) with $f(g)$ and $(2j+1)^{-1/2} c_{lm}^j \Tr \rho\Lambda^j_m$ with $\hat f^j_{m'm}$ (both with fixed $i$), we get the entropic inequalities for (\ref{eq:5.7}).

\section*{Acknowledgments}
VPN's work was supported in part by the U.S. National Science Foundation Grants No. PHY-2112729 and No. PHY-1820271 and by PSC-CUNY awards. FC and AFRL acknowledge financial support from
the Faculty of Sciences of Universidad de los Andes through project INV-2019-84-1833.


\begin{thebibliography}{99}%
\makeatletter
\providecommand \@ifxundefined [1]{%
 \@ifx{#1\undefined}
}%
\providecommand \@ifnum [1]{%
 \ifnum #1\expandafter \@firstoftwo
 \else \expandafter \@secondoftwo
 \fi
}%
\providecommand \@ifx [1]{%
 \ifx #1\expandafter \@firstoftwo
 \else \expandafter \@secondoftwo
 \fi
}%
\providecommand \natexlab [1]{#1}%
\providecommand \enquote  [1]{``#1''}%
\providecommand \bibnamefont  [1]{#1}%
\providecommand \bibfnamefont [1]{#1}%
\providecommand \citenamefont [1]{#1}%
\providecommand \href@noop [0]{\@secondoftwo}%
\providecommand \href [0]{\begingroup \@sanitize@url \@href}%
\providecommand \@href[1]{\@@startlink{#1}\@@href}%
\providecommand \@@href[1]{\endgroup#1\@@endlink}%
\providecommand \@sanitize@url [0]{\catcode `\\12\catcode `\$12\catcode
  `\&12\catcode `\#12\catcode `\^12\catcode `\_12\catcode `\%12\relax}%
\providecommand \@@startlink[1]{}%
\providecommand \@@endlink[0]{}%
\providecommand \url  [0]{\begingroup\@sanitize@url \@url }%
\providecommand \@url [1]{\endgroup\@href {#1}{\urlprefix }}%
\providecommand \urlprefix  [0]{URL }%
\providecommand \Eprint [0]{\href }%
\providecommand \doibase [0]{https://doi.org/}%
\providecommand \selectlanguage [0]{\@gobble}%
\providecommand \bibinfo  [0]{\@secondoftwo}%
\providecommand \bibfield  [0]{\@secondoftwo}%
\providecommand \translation [1]{[#1]}%
\providecommand \BibitemOpen [0]{}%
\providecommand \bibitemStop [0]{}%
\providecommand \bibitemNoStop [0]{.\EOS\space}%
\providecommand \EOS [0]{\spacefactor3000\relax}%
\providecommand \BibitemShut  [1]{\csname bibitem#1\endcsname}%
\let\auto@bib@innerbib\@empty
\bibitem [{\citenamefont {Bia{\l}ynicki-Birula}(1975)}]{BialynickiBirula1975}%
  \BibitemOpen
  \bibfield  {author} {\bibinfo {author} {\bibfnamefont {M.~J.}\ \bibnamefont
  {Bia{\l}ynicki-Birula}, \bibfnamefont {I.}},\ }\bibfield  {title} {\bibinfo
  {title} {Uncertainty relations for information entropy in wave mechanics.},\
  }\href {https://doi.org/https://doi.org/10.1007/BF01608825} {\bibfield
  {journal} {\bibinfo  {journal} {Commun.Math. Phys.}\ }\textbf {\bibinfo
  {volume} {44}},\ \bibinfo {pages} {129} (\bibinfo {year} {1975})}\BibitemShut
  {NoStop}%
\bibitem [{Note1()}]{Note1}%
  \BibitemOpen
  \bibinfo {note} {We use the term ``vector state'' for both a state on
  ${\protect \mathcal A}_S$ and a ray in a background Hilbert space ${\protect
  \mathcal H}$.}\BibitemShut {Stop}%
\bibitem [{\citenamefont {Hepp}(1972)}]{Hepp1972}%
  \BibitemOpen
  \bibfield  {author} {\bibinfo {author} {\bibfnamefont {K.}~\bibnamefont
  {Hepp}},\ }\bibfield  {title} {\bibinfo {title} {{Quantum theory of
  measurement and macroscopic observables}},\ }\href@noop {} {\bibfield
  {journal} {\bibinfo  {journal} {Helv. Phys. Acta}\ }\textbf {\bibinfo
  {volume} {45}},\ \bibinfo {pages} {237} (\bibinfo {year} {1972})}\BibitemShut
  {NoStop}%
\bibitem [{\citenamefont {Araki}\ and\ \citenamefont
  {Yanase}(1960)}]{Araki1960}%
  \BibitemOpen
  \bibfield  {author} {\bibinfo {author} {\bibfnamefont {H.}~\bibnamefont
  {Araki}}\ and\ \bibinfo {author} {\bibfnamefont {M.~M.}\ \bibnamefont
  {Yanase}},\ }\bibfield  {title} {\bibinfo {title} {Measurement of {Q}uantum
  {M}echanical {O}perators},\ }\href {https://doi.org/10.1103/PhysRev.120.622}
  {\bibfield  {journal} {\bibinfo  {journal} {Phys. Rev.}\ }\textbf {\bibinfo
  {volume} {120}},\ \bibinfo {pages} {622} (\bibinfo {year}
  {1960})}\BibitemShut {NoStop}%
\bibitem [{\citenamefont {Fioroni}\ and\ \citenamefont
  {Immirzi}(1994)}]{Fioroni1994}%
  \BibitemOpen
  \bibfield  {author} {\bibinfo {author} {\bibfnamefont {M.}~\bibnamefont
  {Fioroni}}\ and\ \bibinfo {author} {\bibfnamefont {G.}~\bibnamefont
  {Immirzi}},\ }\bibfield  {title} {\bibinfo {title} {How and why the wave
  function collapses after a measurement},\ }\href@noop {} {\bibfield
  {journal} {\bibinfo  {journal} {arXiv preprint gr-qc/9411044}\ } (\bibinfo
  {year} {1994})}\BibitemShut {NoStop}%
\bibitem [{\citenamefont {Wightman}(1995)}]{Wightman1995}%
  \BibitemOpen
  \bibfield  {author} {\bibinfo {author} {\bibfnamefont {A.~S.}\ \bibnamefont
  {Wightman}},\ }\bibfield  {title} {\bibinfo {title} {Superselection rules;
  old and new},\ }\href {https://doi.org/10.1007/BF02741478} {\bibfield
  {journal} {\bibinfo  {journal} {Il Nuovo Cimento B}\ }\textbf {\bibinfo
  {volume} {110}},\ \bibinfo {pages} {751} (\bibinfo {year}
  {1995})}\BibitemShut {NoStop}%
\bibitem [{\citenamefont {Landsman}(2017{\natexlab{a}})}]{Landsman2017a}%
  \BibitemOpen
  \bibfield  {author} {\bibinfo {author} {\bibfnamefont {K.}~\bibnamefont
  {Landsman}},\ }\bibinfo {title} {Niels {B}ohr and the {P}hilosophy of
  {P}hysics: {T}wenty-{F}irst {C}entury {P}erspectives}\ (\bibinfo  {publisher}
  {Bloomsbury},\ \bibinfo {year} {2017})\ Chap.\ \bibinfo {chapter}
  {Bohrification: From classical concepts to commutative algebras}, pp.\
  \bibinfo {pages} {335--366},\ \Eprint {https://arxiv.org/abs/arXiv:1601.02794
  [math-ph]} {arXiv:1601.02794 [math-ph]} \BibitemShut {NoStop}%
\bibitem [{\citenamefont {Landsman}(2017{\natexlab{b}})}]{Landsman2017}%
  \BibitemOpen
  \bibfield  {author} {\bibinfo {author} {\bibfnamefont {K.}~\bibnamefont
  {Landsman}},\ }\href {https://doi.org/10.1007/978-3-319-51777-3} {\emph
  {\bibinfo {title} {Foundations of {Q}uantum {T}heory}}},\ \bibinfo {series}
  {Fundamental Theories of Physics}, Vol.\ \bibinfo {volume} {188}\ (\bibinfo
  {publisher} {Springer Nature},\ \bibinfo {year} {2017})\BibitemShut {NoStop}%
\bibitem [{\citenamefont {Howard}(2007)}]{Howard2007}%
  \BibitemOpen
  \bibfield  {author} {\bibinfo {author} {\bibfnamefont {D.}~\bibnamefont
  {Howard}},\ }\bibfield  {title} {\bibinfo {title} {Revisiting the
  {E}instein--{B}ohr {D}ialogue.},\ }\href
  {https://www.jstor.org/stable/23354465} {\bibfield  {journal} {\bibinfo
  {journal} {Iyyun: The Jerusalem Philosophical Quarterly}\ }\textbf {\bibinfo
  {volume} {56}},\ \bibinfo {pages} {57} (\bibinfo {year} {2007})}\BibitemShut
  {NoStop}%
\bibitem [{\citenamefont {Landsman}(2006)}]{Landsman2006}%
  \BibitemOpen
  \bibfield  {author} {\bibinfo {author} {\bibfnamefont {N.~P.}\ \bibnamefont
  {Landsman}},\ }\bibfield  {title} {\bibinfo {title} {When champions meet:
  {R}ethinking the {B}ohr-{E}instein debate},\ }\href
  {https://doi.org/https://doi.org/10.1016/j.shpsb.2005.10.002} {\bibfield
  {journal} {\bibinfo  {journal} {Studies in History and Philosophy of Science
  Part B: Studies in History and Philosophy of Modern Physics}\ }\textbf
  {\bibinfo {volume} {37}},\ \bibinfo {pages} {212} (\bibinfo {year} {2006})},\
  \bibinfo {note} {2005: The Centenary of Einstein's Annus
  Mirabilis}\BibitemShut {NoStop}%
\bibitem [{\citenamefont {Kadison}\ and\ \citenamefont
  {Singer}(1959)}]{Kadison1959}%
  \BibitemOpen
  \bibfield  {author} {\bibinfo {author} {\bibfnamefont {R.~V.}\ \bibnamefont
  {Kadison}}\ and\ \bibinfo {author} {\bibfnamefont {I.~M.}\ \bibnamefont
  {Singer}},\ }\bibfield  {title} {\bibinfo {title} {Extensions of {P}ure
  {S}tates},\ }\href {https://doi.org/10.2307/2372748} {\bibfield  {journal}
  {\bibinfo  {journal} {American Journal of Mathematics}\ }\textbf {\bibinfo
  {volume} {81}},\ \bibinfo {pages} {383} (\bibinfo {year} {1959})}\BibitemShut
  {NoStop}%
\bibitem [{\citenamefont {Marcus}\ \emph {et~al.}(2015)\citenamefont {Marcus},
  \citenamefont {Spielman},\ and\ \citenamefont {Srivastava}}]{Marcus2015}%
  \BibitemOpen
  \bibfield  {author} {\bibinfo {author} {\bibfnamefont {A.~W.}\ \bibnamefont
  {Marcus}}, \bibinfo {author} {\bibfnamefont {D.~A.}\ \bibnamefont
  {Spielman}},\ and\ \bibinfo {author} {\bibfnamefont {N.}~\bibnamefont
  {Srivastava}},\ }\bibfield  {title} {\bibinfo {title} {Interlacing families
  {II}: {M}ixed characteristic polynomials and the {K}adison-{S}inger
  problem},\ }\href
  {https://doi.org/https://doi.org/10.4007/annals.2015.182.1.8} {\bibfield
  {journal} {\bibinfo  {journal} {Annals of Mathematics}\ }\textbf {\bibinfo
  {volume} {182}},\ \bibinfo {pages} {327} (\bibinfo {year}
  {2015})}\BibitemShut {NoStop}%
\bibitem [{\citenamefont {Marcus}\ and\ \citenamefont
  {Srivastava}(2017)}]{Marcus2017}%
  \BibitemOpen
  \bibfield  {author} {\bibinfo {author} {\bibfnamefont {A.~W.}\ \bibnamefont
  {Marcus}}\ and\ \bibinfo {author} {\bibfnamefont {N.}~\bibnamefont
  {Srivastava}},\ }\bibfield  {title} {\bibinfo {title} {The solution of the
  {K}adison-{S}inger problem},\ }\href@noop {} {\bibfield  {journal} {\bibinfo
  {journal} {arXiv preprint arXiv:1712.08874}\ } (\bibinfo {year}
  {2017})}\BibitemShut {NoStop}%
\bibitem [{\citenamefont {Man'ko}\ and\ \citenamefont
  {Man'ko}(2011)}]{Manko2011}%
  \BibitemOpen
  \bibfield  {author} {\bibinfo {author} {\bibfnamefont {M.~A.}\ \bibnamefont
  {Man'ko}}\ and\ \bibinfo {author} {\bibfnamefont {V.~I.}\ \bibnamefont
  {Man'ko}},\ }\bibfield  {title} {\bibinfo {title} {Dynamic symmetries and
  entropic inequalities in the probability representation of quantum
  mechanics},\ }\href {https://doi.org/10.1063/1.3555483} {\bibfield  {journal}
  {\bibinfo  {journal} {AIP Conference Proceedings}\ }\textbf {\bibinfo
  {volume} {1334}},\ \bibinfo {pages} {217} (\bibinfo {year}
  {2011})}\BibitemShut {NoStop}%
\bibitem [{\citenamefont {Man'ko}\ \emph {et~al.}(2004)\citenamefont {Man'ko},
  \citenamefont {Marmo}, \citenamefont {Sudarshan},\ and\ \citenamefont
  {Zaccaria}}]{Manko2004}%
  \BibitemOpen
  \bibfield  {author} {\bibinfo {author} {\bibfnamefont {V.~I.}\ \bibnamefont
  {Man'ko}}, \bibinfo {author} {\bibfnamefont {G.}~\bibnamefont {Marmo}},
  \bibinfo {author} {\bibfnamefont {E.~C.~G.}\ \bibnamefont {Sudarshan}},\ and\
  \bibinfo {author} {\bibfnamefont {F.}~\bibnamefont {Zaccaria}},\ }\bibfield
  {title} {\bibinfo {title} {Positive maps of density matrix and a tomographic
  criterion of entanglement},\ }\href
  {https://doi.org/https://doi.org/10.1016/j.physleta.2004.05.007} {\bibfield
  {journal} {\bibinfo  {journal} {Physics Letters A}\ }\textbf {\bibinfo
  {volume} {327}},\ \bibinfo {pages} {353} (\bibinfo {year}
  {2004})}\BibitemShut {NoStop}%
\bibitem [{\citenamefont {Ibort}\ \emph {et~al.}(2009)\citenamefont {Ibort},
  \citenamefont {Man{\textquotesingle}ko}, \citenamefont {Marmo}, \citenamefont
  {Simoni},\ and\ \citenamefont {Ventriglia}}]{Ibort2009}%
  \BibitemOpen
  \bibfield  {author} {\bibinfo {author} {\bibfnamefont {A.}~\bibnamefont
  {Ibort}}, \bibinfo {author} {\bibfnamefont {V.~I.}\ \bibnamefont
  {Man{\textquotesingle}ko}}, \bibinfo {author} {\bibfnamefont
  {G.}~\bibnamefont {Marmo}}, \bibinfo {author} {\bibfnamefont
  {A.}~\bibnamefont {Simoni}},\ and\ \bibinfo {author} {\bibfnamefont
  {F.}~\bibnamefont {Ventriglia}},\ }\bibfield  {title} {\bibinfo {title} {An
  introduction to the tomographic picture of quantum mechanics},\ }\href
  {https://doi.org/10.1088/0031-8949/79/06/065013} {\bibfield  {journal}
  {\bibinfo  {journal} {Physica Scripta}\ }\textbf {\bibinfo {volume} {79}},\
  \bibinfo {pages} {065013} (\bibinfo {year} {2009})}\BibitemShut {NoStop}%
\bibitem [{\citenamefont {Connes}\ and\ \citenamefont
  {St{\o}rmer}(1978)}]{Connes1978}%
  \BibitemOpen
  \bibfield  {author} {\bibinfo {author} {\bibfnamefont {A.}~\bibnamefont
  {Connes}}\ and\ \bibinfo {author} {\bibfnamefont {E.}~\bibnamefont
  {St{\o}rmer}},\ }\bibfield  {title} {\bibinfo {title} {Homogeneity of the
  state space of factors of type $\mathrm{III}_1$},\ }\href
  {https://doi.org/https://doi.org/10.1016/0022-1236(78)90085-X} {\bibfield
  {journal} {\bibinfo  {journal} {Journal of Functional Analysis}\ }\textbf
  {\bibinfo {volume} {28}},\ \bibinfo {pages} {187} (\bibinfo {year}
  {1978})}\BibitemShut {NoStop}%
\bibitem [{\citenamefont {Connes}\ \emph {et~al.}(1985)\citenamefont {Connes},
  \citenamefont {Haagerup},\ and\ \citenamefont {St{\o}rmer}}]{Connes1985}%
  \BibitemOpen
  \bibfield  {author} {\bibinfo {author} {\bibfnamefont {A.}~\bibnamefont
  {Connes}}, \bibinfo {author} {\bibfnamefont {U.}~\bibnamefont {Haagerup}},\
  and\ \bibinfo {author} {\bibfnamefont {E.}~\bibnamefont {St{\o}rmer}},\
  }\bibinfo {title} {Operator {A}lgebras and their {C}onnections with
  {T}opology and {E}rgodic {T}heory}\ (\bibinfo  {publisher} {Springer, Berlin,
  Heidelberg},\ \bibinfo {year} {1985})\ Chap.\ \bibinfo {chapter} {Diameters
  of state spaces of type III factors}, pp.\ \bibinfo {pages}
  {91--116}\BibitemShut {NoStop}%
\bibitem [{\citenamefont {O.~E. Lanford~III}(1969)}]{Lanford1969}%
  \BibitemOpen
  \bibfield  {author} {\bibinfo {author} {\bibfnamefont {D.~R.}\ \bibnamefont
  {O.~E. Lanford~III}},\ }\bibfield  {title} {\bibinfo {title} {Observables at
  infinity and states with short range correlations in statistical mechanics},\
  }\href@noop {} {\bibfield  {journal} {\bibinfo  {journal} {Comm. Math.
  Phys.}\ }\textbf {\bibinfo {volume} {13}},\ \bibinfo {pages} {194} (\bibinfo
  {year} {1969})}\BibitemShut {NoStop}%
\bibitem [{Note2()}]{Note2}%
  \BibitemOpen
  \bibinfo {note} {Note that at this step we are already making some preferred
  choice of $\protect \mathcal A_M$. Namely, we start with the diagonal maximal
  abelian subalgebra. This sets up some kind of a ``reference frame'' and all
  other abelian subalgebras will be defined with respect to this choice. See
  also the futher discussion.}\BibitemShut {Stop}%
\bibitem [{\citenamefont {Bapat}\ and\ \citenamefont
  {Raghavan}(1997)}]{Bapat1997}%
  \BibitemOpen
  \bibfield  {author} {\bibinfo {author} {\bibfnamefont {R.~B.}\ \bibnamefont
  {Bapat}}\ and\ \bibinfo {author} {\bibfnamefont {T.~E.~S.}\ \bibnamefont
  {Raghavan}},\ }\href {https://doi.org/10.1017/CBO9780511529979} {\emph
  {\bibinfo {title} {Nonnegative {M}atrices and {A}pplications}}},\ \bibinfo
  {series} {Encyclopedia of Mathematics and its Applications}, Vol.~\bibinfo
  {volume} {64}\ (\bibinfo  {publisher} {Cambridge University Press,
  Cambridge},\ \bibinfo {year} {1997})\BibitemShut {NoStop}%
\bibitem [{\citenamefont {Marcus}\ and\ \citenamefont
  {Ree}(1959)}]{Marcus1959}%
  \BibitemOpen
  \bibfield  {author} {\bibinfo {author} {\bibfnamefont {M.}~\bibnamefont
  {Marcus}}\ and\ \bibinfo {author} {\bibfnamefont {R.}~\bibnamefont {Ree}},\
  }\bibfield  {title} {\bibinfo {title} {Diagonals of {D}oubly {S}tochastic
  {M}atrices},\ }\href {https://doi.org/10.1093/qmath/10.1.296} {\bibfield
  {journal} {\bibinfo  {journal} {The Quarterly Journal of Mathematics}\
  }\textbf {\bibinfo {volume} {10}},\ \bibinfo {pages} {296} (\bibinfo {year}
  {1959})},\ \Eprint
  {https://arxiv.org/abs/https://academic.oup.com/qjmath/article-pdf/10/1/296/7288034/10-1-296.pdf}
  {https://academic.oup.com/qjmath/article-pdf/10/1/296/7288034/10-1-296.pdf}
  \BibitemShut {NoStop}%
\bibitem [{\citenamefont {Sorkin}(1993)}]{Sorkin1993}%
  \BibitemOpen
  \bibfield  {author} {\bibinfo {author} {\bibfnamefont {R.}~\bibnamefont
  {Sorkin}},\ }\bibfield  {title} {\bibinfo {title} {Impossible {M}easurements
  on {Q}uantum {F}ields},\ }in\ \href@noop {} {\emph {\bibinfo {booktitle}
  {Directions in General Relativity: Proceedings of the 1993 International
  Symposium, Maryland}}},\ Vol.~\bibinfo {volume} {2},\ \bibinfo {editor}
  {edited by\ \bibinfo {editor} {\bibfnamefont {B.-L.}\ \bibnamefont {Hu}}\
  and\ \bibinfo {editor} {\bibfnamefont {T.~A.}\ \bibnamefont {Jacobson}}}\
  (\bibinfo  {publisher} {Cambridge U.P.},\ \bibinfo {year} {1993})\ \Eprint
  {https://arxiv.org/abs/arXiv:gr-qc/9302018v2} {arXiv:gr-qc/9302018v2}
  \BibitemShut {NoStop}%
\bibitem [{\citenamefont {Reed}\ and\ \citenamefont {Simon}(1975)}]{Reed1975}%
  \BibitemOpen
  \bibfield  {author} {\bibinfo {author} {\bibfnamefont {M.}~\bibnamefont
  {Reed}}\ and\ \bibinfo {author} {\bibfnamefont {B.}~\bibnamefont {Simon}},\
  }\href@noop {} {\emph {\bibinfo {title} {Fourier {A}nalysis,
  {S}elf-{A}djointness}}},\ \bibinfo {series} {Methods of Modern Mathematical
  Physics}, Vol.~\bibinfo {volume} {2}\ (\bibinfo  {publisher} {Academic
  Press},\ \bibinfo {year} {1975})\BibitemShut {NoStop}%
\bibitem [{\citenamefont {Babenko}(1961)}]{Babenko1961}%
  \BibitemOpen
  \bibfield  {author} {\bibinfo {author} {\bibfnamefont {K.~I.}\ \bibnamefont
  {Babenko}},\ }\bibfield  {title} {\bibinfo {title} {An inequality in the
  theory of {F}ourier integrals},\ }\href@noop {} {\bibfield  {journal}
  {\bibinfo  {journal} {Izv. Akad. Nauk SSSR Ser. Mat.}\ }\textbf {\bibinfo
  {volume} {25}},\ \bibinfo {pages} {531} (\bibinfo {year} {1961})}\BibitemShut
  {NoStop}%
\bibitem [{\citenamefont {Beckner}(1975)}]{Beckner1975}%
  \BibitemOpen
  \bibfield  {author} {\bibinfo {author} {\bibfnamefont {W.}~\bibnamefont
  {Beckner}},\ }\bibfield  {title} {\bibinfo {title} {Inequalities in {F}ourier
  {A}nalysis},\ }\href {http://www.jstor.org/stable/1970980} {\bibfield
  {journal} {\bibinfo  {journal} {Annals of Mathematics}\ }\textbf {\bibinfo
  {volume} {102}},\ \bibinfo {pages} {159} (\bibinfo {year}
  {1975})}\BibitemShut {NoStop}%
\bibitem [{\citenamefont {Cowling}\ \emph {et~al.}(2019)\citenamefont
  {Cowling}, \citenamefont {Martini}, \citenamefont {M{\"u}ller},\ and\
  \citenamefont {Parcet}}]{Cowling2019}%
  \BibitemOpen
  \bibfield  {author} {\bibinfo {author} {\bibfnamefont {M.~G.}\ \bibnamefont
  {Cowling}}, \bibinfo {author} {\bibfnamefont {A.}~\bibnamefont {Martini}},
  \bibinfo {author} {\bibfnamefont {D.}~\bibnamefont {M{\"u}ller}},\ and\
  \bibinfo {author} {\bibfnamefont {J.}~\bibnamefont {Parcet}},\ }\bibfield
  {title} {\bibinfo {title} {The {H}ausdorff--{Y}oung inequality on {L}ie
  groups},\ }\href@noop {} {\bibfield  {journal} {\bibinfo  {journal}
  {Mathematische Annalen}\ }\textbf {\bibinfo {volume} {375}},\ \bibinfo
  {pages} {93} (\bibinfo {year} {2019})}\BibitemShut {NoStop}%
\end{thebibliography}
%
\end{document}